\documentclass{cpbtex}
\usepackage{txfonts}
\usepackage{mathrsfs}
\usepackage[numbers]{natbib}
\usepackage{siunitx}
\usepackage{graphicx}
\begin{document}

\def\HI{H\,{\small{I}} }
\newcommand{\kms}{\mathrm{km\,s^{-1}}}
\newcommand{\citetn}[1]{\citeauthor{#1}~\cite{#1}}
\makeatletter
\newcommand{\ancite}[1]{%
  \@ifundefined{r@#1}{{\bfseries ??}}{%
    \expandafter\expandafter\expandafter
    \@ancite\csname r@#1\endcsname{#1}%
  }%
}

\def\@ancite#1#2{%
  \begingroup
  \def\@cite##1##2{[##1\if@tempswa , ##2\fi]}%
  \let\citeauthoryear\@citeauthoryear
  \let\citeN\@citeN
  \@citeauthor{#2}~\cite{#2}%
  \endgroup
}
\makeatother

\title{HiFAST: An \HI data calibration and imaging pipeline for the FAST IV: The stray-radiation correction}


\author{
Qing-ze Chen (陈箐泽) $^{1,3}$\thanks{Corresponding author. E-mail:chenqz@bao.ac.cn}, \ 
Jie Wang (王杰) $^{1,2,3}$, \
Ying-jie Jing (景英杰) $^{1}$, \\
Li-gang Hou (侯立刚) $^{1}$, \
Chen Xu (徐晨) $^{1,3}$, \
Tian-tian Liang (梁甜甜) $^{1,3}$, \\
Xu-yang Gao (高旭阳) $^{1,4}$, \
Jin-lin Han (韩金林) $^{1,4}$, \
Zi-ming Liu (刘孜铭) $^{1}$, \\
Bin Liu (刘彬) $^{1}$, \
Chuan-peng Zhang (张传朋) $^{1}$, \
Heng-qian Gan (甘恒谦) $^{1}$, \\
Ming Zhu (朱明) $^{1}$, \
Yan Zhu (朱岩) $^{1}$, \
and \ Peng Jiang (姜鹏) $^{1,4}$\\
$^{1}$National Astronomical Observatories, Chinese Academy of Sciences, Beijing 100012, China\\
$^{2}$Institute for Frontiers in Astronomy and Astrophysics, Beijing Normal University,\\ Beijing 102206, China\\
$^{3}$School of Astronomy and Space Science, University of Chinese Academy of Sciences,\\ Beijing 100049, China\\
$^{4}$CAS Key Laboratory of FAST, National Astronomical Observatories,\\  Chinese Academy of Sciences, Beijing 100101, China
}


\date{\today}
\maketitle

\begin{abstract}
Stray radiation is a considerable challenge for radio telescopes, requiring careful assessment due to its effects. This is crucial when the strong background flux from side lobes significantly affects the total flux, especially for extended sources. In this study, we introduced the beam pattern of the L-band receiver on the Five-hundred-meter Aperture Spherical Telescope (FAST), covering various frequencies based on recent observations. We discovered that the main beam efficiency of all beams exceeds 90\% throughout the L band frequencies, with efficiency decreasing slowly as frequency increases. Subsequently, we developed a module to mitigate stray radiation effects, incorporating it into FAST's standard \HI data reduction process, referred to as \texttt{HiFAST}. Our analysis shows that side lobe flux's influence, particularly for extended sources with significant surface density gradients, necessitates detailed evaluation. Corrections for the extended M33 galaxy can reach up to 20\%. Moreover, the pattern data presented here is vital for studying HI intensity maps at high redshift. The module, along with HiFAST and beam pattern data across 15 frequency bins, can be accessed at \textrm{https://hifast.readthedocs.io}. The datasets of beam pattern presented in this paper, are openly available at \textrm{https://doi.org/10.57760/sciencedb.j00113.00266} (https://www.scidb.cn/s/bqQRNv).
\end{abstract}

\begin{center}

\end{center}

\section{Introduction}
A single-dish radio telescope responds to a point source as a point spread function (PSF). The ideal PSF is an Airy disc featuring a central maximum and diffraction rings, referred to as the main beam and sidelobes in a radio telescope. In practice, the side lobes of a real telescope are higher due to obstructions from the feed cabin and the supporting legs on the reflector. Emissions entering the receiver via the antenna's side lobes are considered stray radiation; thus, side lobes are also called stray patterns.

At present, FAST is equipped with a 19-beam L-band receiver. The beam positions in the focal plane differ from one another, and each beam has its own distinct beam pattern. When a source is observed with multiple beams, the final spectrum is obtained by integrating the signals from all these beams. This makes correcting for stray radiation a big challenge.

FAST uses six steel cables to support the feed cabin, which has a diameter of 13 m \citep{Jiang2019}.
The cable-supporting structure and the relatively smaller size of the feed cabin reduced the shielding effect on the reflector. With the reduced shielding effect, the stray radiation of FAST has diminished, and a higher main beam efficiency can be achieved.

At 3 GHz, the simulation of the FAST beam pattern \citep{Gan2020} indicates that the side lobes are 40 dB weaker compared to the peak of the main beam when the bias angle surpasses 0.1$^\circ$. For our analysis, we applied the GRASP package to simulate the beam pattern at 1.4 GHz. At this lower frequency, the beam size increases, and the side lobes are approximately 30 dB weaker than the peak of the main beam when measured at 10 arc minutes. Test results suggest \citep{Jiang_2020} that beyond 10 arcminutes from the source, the stray pattern is not detectable. Excluding far side lobes satisfies accuracy requirements, as their stray radiation is negligible compared to near side lobes. Nevertheless, it remains crucial to assess the distribution of the main and side lobes for various beams and across diverse frequency bands using actual observational data. This paper aims to provide vital and foundational information for all users.

Moreover, while the side lobes do not significantly impact the functionality of the FAST telescope, stray radiation remains a considerable threat when observing the \HI 21 cm spectral line. This concern arises from the widespread and strong \HI emission of the Milky Way, which enters through the side lobes and remains noticeable for the observation of local sources. Previous studies on \HI emissions from the Milky Way have implemented stray radiation correction methods \citep{1980AA....82..275K,Hartmann1996,Kalberla2005,Kalberla2010} due to the pervasive nature of Milky Way \HI emissions originating from all areas of the sky. In addition to diffuse \HI emissions from the Milky Way covering the entire celestial expanse, extended sources that can be spatially resolved by FAST are susceptible to the effects of stray radiation as well. For extended sources, the distribution of radiation intensity within the source is altered, since stray radiation at any given location stems from other locations. Our stray radiation correction technique is designed to mitigate the impact.

Kalberla et al. \citep{1980AA....82..275K} first found a method to obtain true brightness temperature from the measured antenna temperature, mixed with stray radiation. The method estimates the fraction of stray radiation for a point and subtracts that fraction. Subsequently, stray radiation corrections were implemented by Hartmann et al. \citep{Hartmann1996} and Kalberla et al. \citep{Kalberla2005} on the Dwingeloo telescope, by Higgs et al. \citep{Higgs2005} on the 26-metre telescope at the Dominion Radio Astrophysical Observatory, by Bajaja et al. \citep{Bajaja2005} on the 30-metre telescope at Villa Elisa, by Kalberla et al. \citep{Kalberla2010} on the Parkes telescope, by Boothroyd et al. \citep{Boothroyd2011} on the Green Bank Telescope, and by Winkel et al. \citep{Winkel2016} on the Effelsberg telescope. These efforts achieved a more accurate mapping of the Milky Way's neutral hydrogen through the application of stray radiation correction, a method that can also be applied to study extra-galactic objects.

Xu et al. \citep{Xu2022} observed the \HI of Stephan’s Quintet with FAST and detected the surrounding \HI gas of the galaxy group. Stray radiation correction was applied to the observation to demonstrate that the signal originates from the surrounding gas and not from the artificial signal of the stray radiation from the galaxies. They deconvolved the \HI map with the main beam pattern of the central beam of the FAST 19-beam receiver and estimated the fraction of stray radiation. Their estimation of the stray radiation is especially applied on tracking observation mode. 

In this study, we aim to evaluate the patterns of the main and side lobes of the L-band receiver on FAST and employ the algorithm originally developed by Kalberla et al. \citep{1980AA....82..275K} to adjust for stray radiation. These results will enhance a module within the comprehensive \HI pipeline, \texttt{HiFAST} \footnote{https://hifast.readthedocs.io} , which serves as an \HI data calibration and imaging tool for FAST \citep{Jing_2024}. This pipeline involves frequency-dependent noise diode calibration, band-pass calibration, baseline fitting, standing wave removal \citep{Xu_2025}, as well as flux and gain-curve calibration \citep{Liu_2024}, culminating in stray radiation correction and the gridding necessary to generate a data cube.

This paper is organised as follows. In Section \ref{sec:method} , we begin by presenting the observational data and methodological framework. Section \ref{sec:results}\; details our results, while Section \ref{sec:CD}\;  provides the discussion and conclusions.

\section{Observations and Method} \label{sec:method}

\subsection{Observational Data}

The data used in this paper include the scan data of the strong continuum point source to measure the beam patterns of the 19 beam receivers and the data to evaluate the performance of our algorithm, examining both the accuracy and the effect of stray radiation correction on extended sources.

To map the beam pattern, we observed the radio galaxy PKS 0531+19 on March 7, 19, 20, and 21, 2023, using OnTheFlyMapping mode with a scanning speed of 15 arcminutes per minute and a scan gap of 1 arcminute.
To achieve a high spatial sampling rate, the four observations were performed with scanning paths offset by 15 arcseconds from one another.
We applied our correction algorithm using observations of M33, conducted in drift mode from November 21, 2021, to February 6, 2022. During these observations of M33, the rotation angle and sampling time were adjusted to adhere to the Nyquist sampling criterion.

\begin{center}
\includegraphics[width=16.0cm]{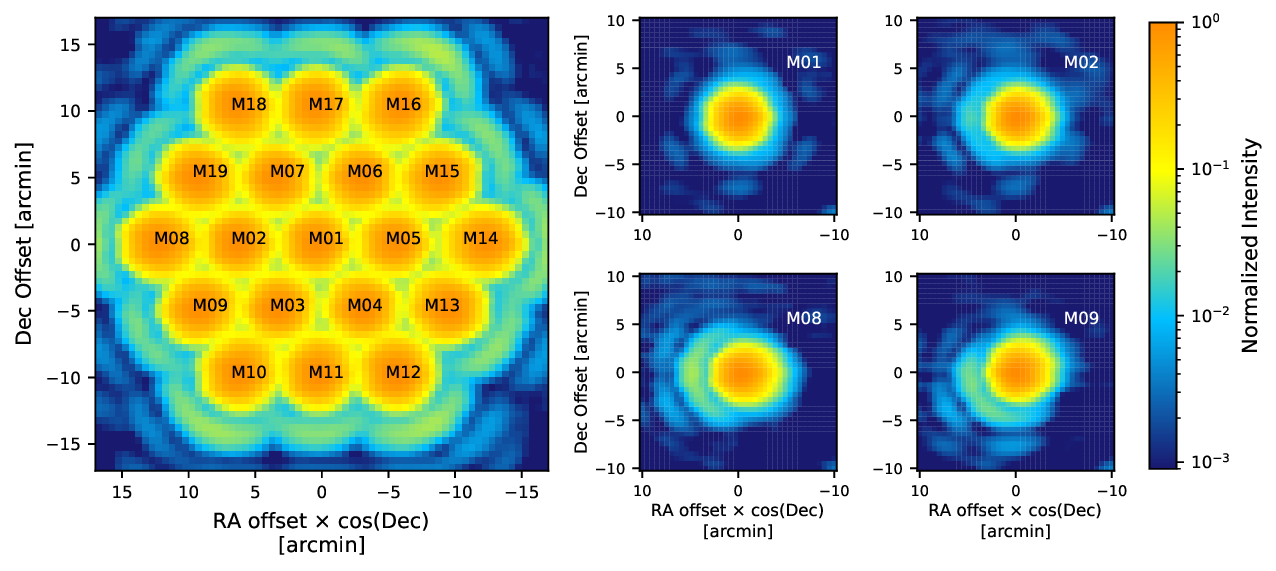}\\[5pt]
\parbox[c]{15.0cm}{\footnotesize{\bf Fig.~1.}   \textbf{Left}: The configuration of FAST 19 beams.
\textbf{Right}: The beam pattern of 4 beams at the frequency band from 1390 MHz to 1410 MHz. M01 is the central beam. M02 is located on the first ring.
M09 and M08 are on the second and the third rings with  $d_{M01}$ of arcmin 10.0 and 11.6 arcmin respectively.
The beam patterns are obtained from the observation of PKS 0531+19.}
\end{center}


\subsection{Beams patterns of FAST 19-beam receiver}\label{sec:beam}

Operating in the 1050-1450 MHz band, the L-band 19-beam receiver provides coverage of the 21 cm \HI emission line. We divide the FAST 19 beams into four rings in the following discussion. Beam M01 is positioned centrally; beams M02 to M07 are located on the first ring, and beams M08 to M19 form the second and third rings. Beams in the first ring are approximately 5.8 arcmin from the central beam, while those in the second and
third rings are roughly 10.0 and 11.6 arcmin away, respectively. Radio telescopes detect signals based on the point spread function (PSF), which is influenced by the telescope's physical configuration. Jiang et al. \citep{Jiang_2020} characterised the pattern, including the main and side lobes, for the 19-beam receiver. We have refined the beam pattern using new observational data. The beam pattern was analysed in two frequency ranges: 1050–1110\,MHz and 1330–1430\,MHz. Especially, Figure 1 illustrates the spatial layout of all 19 beams and their patterns at four distinct distances from the central beam at the at the frequency band from 1390
MHz to 1410 MHz. The central beam M01 features the lowest sidelobe level.
Because the beam pattern and efficiency vary with position in the focal-plane array, M01 denotes the central beam. We define $d_{M01}$ as the angular separation between a given beam and the central beam M01 on the sky.
As $d_{M01}$ increases, the asymmetrical coma aberration in the main beam becomes more pronounced, and the side lobe flux fraction also rises.

Typically, the antenna pattern $P(x,y)$ is determined by observing a radio source, and this pattern emerges from convolving the source with the beam pattern. When the source's angular size is much smaller than that of the beam, the resultant convolved pattern closely resembles the beam pattern itself. Consequently, it is critical that the source used for beam pattern measurement meets specific criteria. The source should have a sufficiently small size. The perceived beam pattern is a convolution of the beam's PSF and the source, and a smaller source size results in a convoluted outcome that closely mirrors the PSF. Moreover, the source's flux should be ample enough to render the sidelobe distinguishable. The source must also be isolated, as other sources within 10 arcminutes could affect the detected sidelobe. We used data from NVSS \citep{1998AJ....115.1693C} to assist in selecting the source. Through this data, we can evaluate the source's flux and environment. At 1.4 GHz, the NVSS cube data reveal a FWHM beam size of 45 arcseconds, roughly a quarter of the size of the FAST beam. Therefore, if a source is not resolved in the NVSS cube data, its size will be considerably smaller than both the NVSS and FAST beam sizes. According to optical observations \citep{2003A&A...412...45P}, the source we chose, PKS 0531+19, has a major axis of 0.34 arcminutes and a minor axis of 0.22 arcminutes.

The mapping was carried out using on-the-flying scanning. The row spacing (15 arcseconds) and the along-scan sampling interval (7.5 arcseconds) were both set to be well below the Nyquist sampling requirement ($\simeq$ 1.5 arcminutes), ensuring sufficient spatial oversampling of the beam. After baseline subtraction to remove instrumental and sky background contributions, the calibrated data were gridded into two-dimensional maps and normalized to the peak response to obtain the antenna pattern $P(x,y)$.

In Figure 2, we present the beam patterns for four beams across four frequency bands: beam M01 is positioned centrally, beam M02 lies in the first ring, and beams M09 and M08 are situated in the second and third rings. Beam M01 exhibits a nearly symmetrical pattern; however, the main- and side-lobe patterns of the other three beams are slightly off-centre, leading to some asymmetry. The patterns also vary with frequency, as at lower frequencies the flux from the side lobes becomes slightly stronger.
In addition, the beam size shows a frequency dependence, increasing slightly toward lower frequencies as the wavelength becomes longer. This effect should be taken into account when intensity maps are used to derive the variation of the H I fraction with redshift, for which beam patterns at the corresponding frequencies are required.
Here, only four beams over four frequencies are presented for illustration. The beam patterns are measured for all beams, covering two frequency ranges: 1050–1110 MHz and 1330–1430 MHz. All the datasets of beam patterns presented in this paper, are compiled in the Supplementary Materials in FITS format.

\begin{center}
\includegraphics[width=16.0cm]{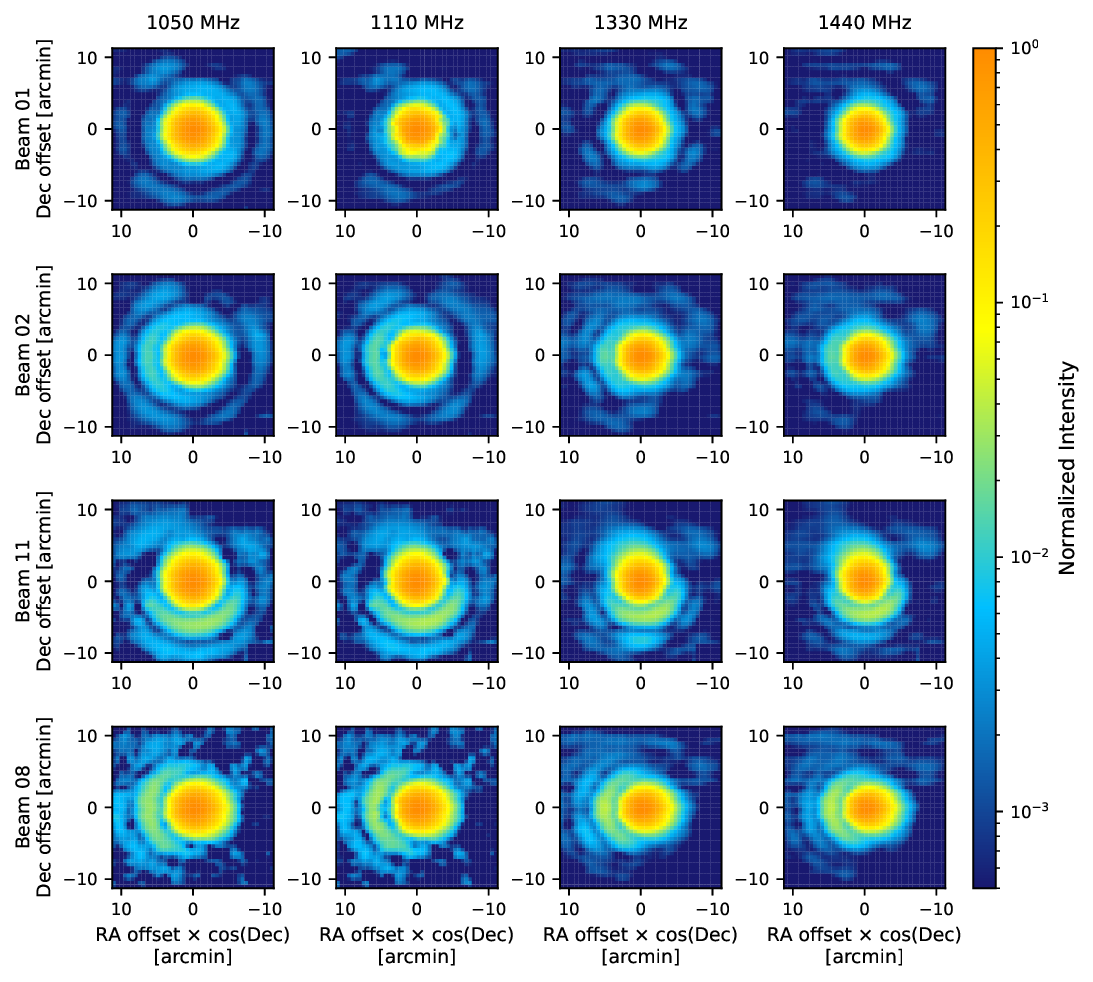}\\[8pt]
\parbox[c]{15.0cm}{\footnotesize{\bf Fig.~2.}   
Normalized beam patterns of Beams 01, 02, 11, and 08 across four frequency bands. The patterns are centered at 1050, 1110, 1330, and 1440 MHz, each integrated over a 20 MHz bandwidth. These beams are located at different rings with angular offsets ($d_{\mathrm{M01}}$) of 0, 5.8, 10.0, and 11.6 arcmin, respectively. The color scale represents the signal intensity, normalized to the peak value of each panel.}
\end{center}

\subsection{Correction Algorithm} \label{Cor}

Kalberla et al. \citep{1980AA....82..275K} introduced a technique aimed at minimising stray radiation. This stray radiation algorithm has been implemented in various single-dish telescopes \citep{Hartmann1996, Kalberla2005, Kalberla2010}. We employ this algorithm to adjust for stray radiation in the FAST single dish. The antenna captures brightness temperature $T_a(x,y)$ from the sources. The expression for $T_a(x,y)$ is the convolution of the actual brightness temperature $T_b(x,y)$ and the antenna pattern $P(x,y)$, represented as
\begin{equation}
T_a(x,y)=\int{T_b(x',y')}P(x-x',y-y')\mathrm{d}x'\mathrm{d}y'.
\end{equation}

When the source's angular size is smaller than the beam size, the observed convoluted pattern will resemble the beam pattern. The antenna pattern $P(x,y)$ can be obtained from beam-mapping observations of a strong, compact flux calibrator. The detailed observational implementation and data-reduction procedure adopted in this work are described in Sect.~2.2.
It is normalized such that $\iint P(x,y) \mathrm{d}x\mathrm{d}y = 1$. Normally, $P(x,y)$ can be divided into two spatial regions: the main beam (MB) and the stray pattern (SP) components, given by $\int P(x,y)\mathrm{d}x\mathrm{d}y=\int_{(mb)} P(x,y)\mathrm{d}x\mathrm{d}y+\int_{(sp)} P(x,y)\mathrm{d}x\mathrm{d}y$. The boundary between the main beam and the stray pattern is determined by the position of the first minimum from the centre outward. With the beam pattern, by area, it can be segregated into the main beam and side lobes, enabling the equation to be reformulated as

\begin{equation}
\begin{array}{ll}
     T_a(x,y)&=\int_{(mb)}{P(x-x',y-y')T_b(x',y')\mathrm{d}x'\mathrm{d}y'}\\
     &+\int_{(sp)}{P(x-x',y-y')T_b(x',y')\mathrm{d}x'\mathrm{d}y'}.
\end{array}
\end{equation}

The main beam efficiency could then be defined as 
\begin{equation}
\begin{array}{ll}
     \eta_{mb}=\int_{(mb)}{P(x,y)\mathrm{d}x\mathrm{d}y}. 
\end{array}
\end{equation}
Since the beam pattern is normalised to unity, the main beam efficiency, as defined here, is less than one.
The main beam efficiencies of the 19 beams are measured in the frequency range from 1400 MHz to 1420
MHz, based on our observations, as shown in Figure 3.
The $\eta_{mb}$ of the central beam M01 can reach 0.97, and the $\eta_{mb}$ of beams located on the outermost rings is around 0.92. 
We also measured the change of main beam efficiency with frequency, which is shown in Figure 4.
\begin{center}
\includegraphics[width=7.0cm]{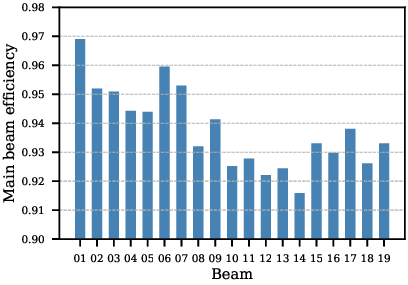}\\[5pt]
\parbox[c]{15.0cm}{\footnotesize{\bf Fig.~3.}   Main beam efficiency of 19 beams measured at the frequency range from 1400 MHz to 1420 MHz.}
\end{center}

\begin{center}
\includegraphics[width=16.0cm]{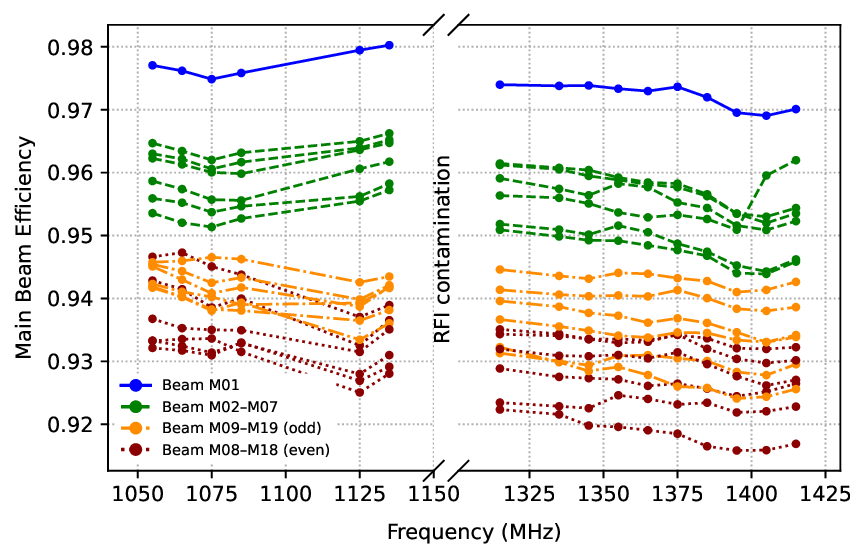}\\[5pt]
\parbox[c]{15.0cm}{\footnotesize{\bf Fig.~4.}  The figure presents the main beam efficiency ($\eta_{\mathrm{MB}}$) of the 19 beams (M01–M19) as a function of frequency. The beams are color-coded as follows: M01 is shown in blue; M02–M07 in green; beams on the second ring (M08–M19, odd-numbered) in orange; and beams on the third ring (M08–M19, even-numbered) in red. The x-axis is discontinuous to exclude the frequency range affected by radio frequency interference (RFI).}
\end{center}

Since the telescope cannot resolve fluctuations in the brightness temperature within its beam, 
it is reasonable to assume that the spatial variation of $T_b$ across the beam is small. 
Consequently, the integral 
\(
    \int_{(mb)} P(x - x', y - y')\, T_b(x', y')\, \mathrm{d}x'\, \mathrm{d}y',
\)
which represents the convolution of the beam pattern with the brightness temperature distribution, 
can be approximated by the product 
$\eta_{mb}\, \overline{T_b}(x, y)$,
reflecting the fraction of the brightness temperature detected within the main beam. 
Accordingly, the expression for $T_a$ can be rewritten as
\begin{equation}
\begin{aligned}
    T_a(x, y) ={}& \eta_{mb}\, \overline{T_b}(x, y) \\
    &+ \int_{(sp)} P(x - x', y - y')\, T_b(x', y')\, \mathrm{d}x'\, \mathrm{d}y',
\end{aligned}
\label{eq:Ta_expression}
\end{equation}
where the second term accounts for the contribution from the side lobes.

Consequently, the formulation for the true brightness temperature of the source, denoted as $T_b(x,y)$, can be articulated as follows:

\begin{equation}\label{eq:Tb}
\begin{array}{ll}
    &\overline{T_b}(x,y)=\frac{T_a(x,y)}{\eta_{mb}}\\
    &-\frac{1}{\eta_{mb}}\int_{(sp)}{P(x-x',y-y')T_b(x',y')\mathrm{d}x'\mathrm{d}y'}.
\end{array}
\end{equation}

In equation~\ref{eq:Tb}, the contribution from the sidelobes is expressed as a spatial integral over 
$T_b(x,y)$ weighted by the stray pattern. 
$\overline{T_b}(x,y)$ can be interpreted as the average brightness temperature within the main beam.
$\overline{T_b}$ can be regarded as ${T_b}$ since we cannot resolve any details of $T_b$ within the beam size.
By substituting $\overline{T_b}(x,y)$ with $T_b(x,y)$, the equation transforms into a Fredholm integral equation. This transformation enables the equation to be solvable when the condition $\eta_{mb}>0.5$ is met, unleashing its potential for analysis.

\begin{equation}
\label{eq-1}
\begin{array}{ll}
    &T_b(x,y)=\frac{T_a(x,y)}{\eta_{mb}}\\
    &-\frac{1}{\eta_{mb}}\int_{(sp)}{P(x-x',y-y')T_b(x',y')\mathrm{d}x'\mathrm{d}y'}
\end{array}
\end{equation}

In equation~\ref{eq-1}, both sides contain the term $T_b$. The $T_b(x, y)$ on the left-hand side represents the brightness temperature  after the stray radiation correction, which is the quantity we aim to derive. In contrast, $T_b(x', y')$ on the right-hand side refers to the  input sky brightness temperature, which remains unknown at this stage.
Typically, we substitute the unknown input sky brightness temperature $T_b(x',y')$ with the known antenna temperature $T_a(x',y')$. This swap has a minimal effect on the correction accuracy since the difference between $T_a(x',y')$ and $T_b(x',y')$ is minor. Though solving equation \ref{eq-1} precisely can be done through iteration, in our scenario where $\eta_{mb}>0.9$, an accurate $T_b(x,y)$ estimate does not require repeated iterations; we will delve into this in section \ref{sec:results} . Hartmann et al. \citep{Hartmann1996} outlines a method using the resolve-kernel derived from the beam pattern. Mathematically, iteration results equate to using the resolve-kernel when dealing with a single beam.

In a multi-beam system, each beam exhibits distinct characteristics in terms of pattern shape and side-lobe orientation, as described in Section~\ref{sec:beam} . During observations, these beams cover different regions of the sky, and the data from each beam are recorded separately. The signals received by each beam contain both direct and stray radiation components. To correct for stray radiation, we constructed an input sky map that incorporates the antenna pattern of all beams through a gridding process. Stray radiation correction is then applied individually to the data from each beam by comparing the input sky map with the corresponding beam pattern. The corrected data from all beams are subsequently combined into a single \texttt{FITS} cube using the \texttt{HiFAST} pipeline. If higher precision is required, the newly generated sky map can serve as the input brightness temperature for another iteration of correction.

\label{subsec:data}

\section{Results} \label{sec:results}
In this section, we will explain how correcting stray radiation affects observational results.

We evaluated the impact of stray radiation correction on typical point sources, such as galaxies or other compact objects whose angular diameters are smaller than the beam size of the FAST. 
Stray radiation tends to scatter a portion of the incoming signal, resulting in a slight reduction in the main beam intensity within the region of interest. 
These sources usually do not create stray radiation that is distinguishable from noise. And if there is no strong background radiation nearby, and the flux from the side lobes is very weak, then the correlation of the stray radiation is insignificant compared with the background noise, and we can pick up a small signal from the radiation. 

The compactness of stray radiation correction is significant for extended sources. Typically, the observed flux from a specific sky pointing is a combination of the main beam's flux at its closest locations and contributions from the sidelobes of nearby sources or background radiation. Correcting the main beam efficiency will lead to an increase in flux, whereas adjusting for the side lobes will reduce the total flux. When examining points across the sky, if the average flux in a nearby area is lower than the flux of the target point, the main beam frequency correction  will dominate, resulting in an increased corrected flux. Conversely, if the nearby average flux is higher, the sidelobe correction prevails, and the total flux decreases. Figure 5 presents the flux distribution in the form of a contour map with colour-coded stray radiation corrections for the M33 region at a velocity of $-100 \, \mathrm{km \, s^{-1}}$. Additionally, the zeroth moments before and after the correction, as well as their difference, are depicted. Across the entire extended source, we observed an increase in flux ranging from a few percent to 10 percent. However, in areas with steep gradients near the peak, where side lobe contributions from the peak are significant, the total flux decreases by several percent, and in some cases, by more than 10 percent. These findings underscore the importance of stray radiation correction for extended sources, particularly in regions close to a strong \HI region.

Figure 6 illustrates the pixel flux in the zeroth moment of M33 both before and after correcting for stray radiation. Most pixels experienced an increase in flux post-correction; however, some showed a decrease. The pixels aligned along the line exhibit a slope marginally exceeding $1$, approximately $1.06$. A linear fit was applied to the data, resulting in a slope of 1.034.

\begin{center}
\label{fig:ext} 
\includegraphics[width=16.0cm]{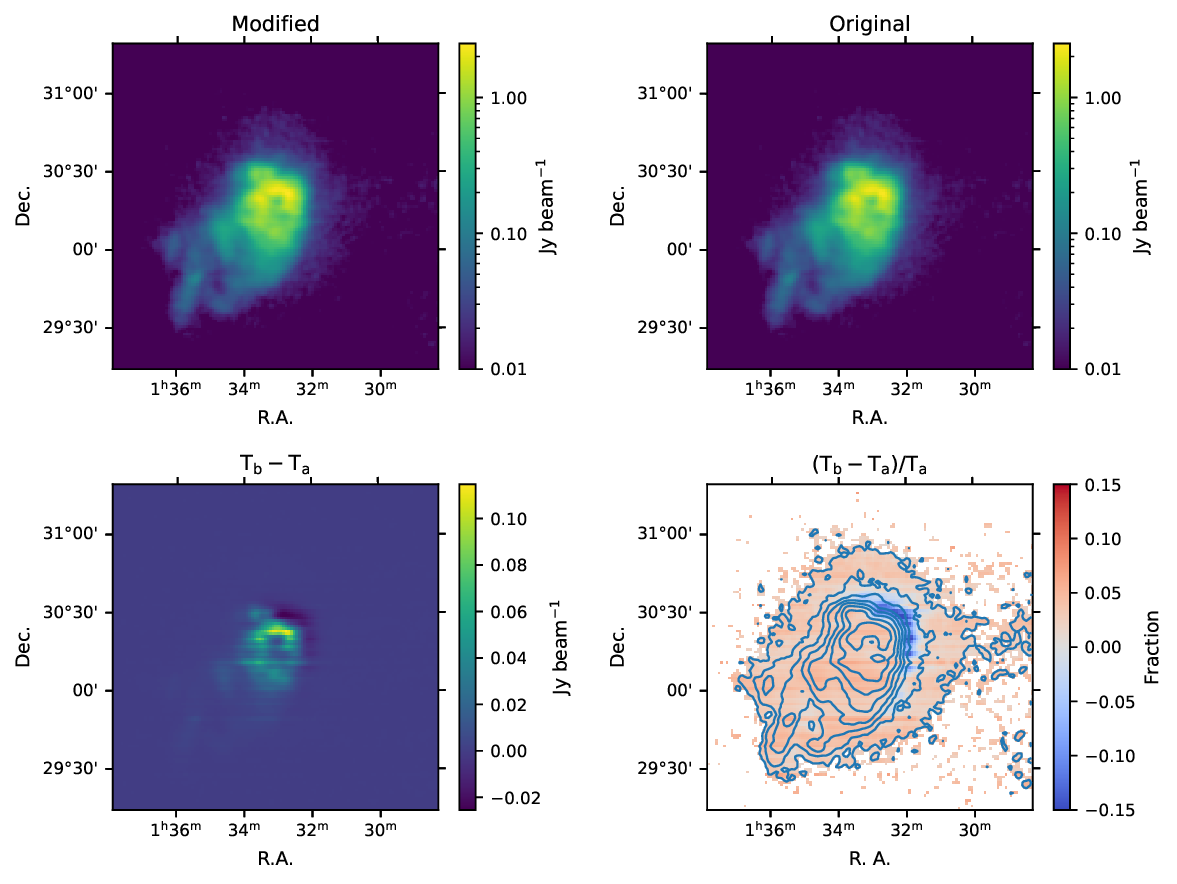}\\[8pt]
\parbox[c]{15.0cm}{\footnotesize{\bf Fig.~5.}   Figure of stray radiation correction effects on the observational data of M33 at the barycentric velocity of $- 100\ km\cdot s^{-1}$.
\textbf{Top left}: Modified data after stray radiation correction .
\textbf{Top right}: Original data.
\textbf{Bottom left}: Difference between modified and original data.
\textbf{Bottom right}: The fraction of stray radiation correction.
Contours represent flux levels at 0.01, 0.02, 0.04, 0.08, 0.16, 0.32, 0.64, and 1.28\,Jy\,beam$^{-1}$.
The color map shows the ratio of flux increment and decrement after stray radiation correction.}
\end{center}

\begin{center}
\label{fig:pix}
\includegraphics[width=7.0cm]{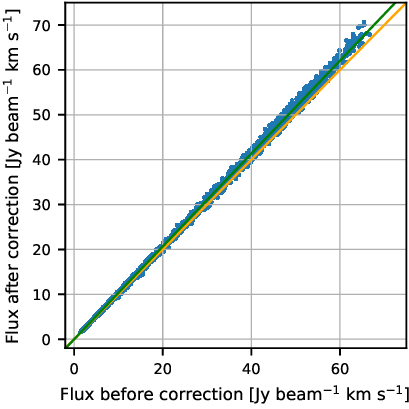}\\[5pt]
\parbox[c]{15.0cm}{\footnotesize{\bf Fig.~6.}   Comparison between the flux of each pixel in the zeroth moment of M33.
The X axis represents flux before correction.
The Y axis represents flux after correction.
The yellow line represents X=Y.
Dots represent the flux in each pixel before and
after correction.
The green line represents its linear fit.}
\end{center}

When solving the equation \ref{eq-1} using iterative methods, the number of iterations influences the accuracy of the solution. Generally, increasing iterations enhances the correction's precision but also requires more time and may yield only minor improvements. To examine this effect, we utilised the M33 observational data. Let $T_a$ represent the initial antenna temperature and $T_b$ the sky's brightness temperature. The corrected temperature is indicated as $T_n$, where $n$ denotes the iteration count. For the initial correction, $T_a$ was used as the input sky brightness temperature $T_b$, resulting in the brightness temperature $T_1$. This $T_1$ is closer to the required brightness temperature $T_b$ than $T_a$, as it being the outcome of stray radiation correction. We subsequently used $T_1$ to obtain $T_2$ and continued similarly for $T_3$. The process was halted at $T_3$ as the difference between $T_2$ and $T_3$ became much smaller than  the difference between $T_1$ and $T_2$. We evaluated the spectra of adjusted $T_1$ and $T_3$ against the original $T_a$ spectrum at a point containing both the M33 and Milky Way signals.
We calculated the the root-mean-square (RMS) of the spectral differences within the velocity range $-200~\kms$ to $20~\kms$.
We obtained
$\mathrm{RMS}(T_a-T_3)=3.25\times10^{-3}$Jy beam$^{-1}$ and $\mathrm{RMS}(T_a-T_1)=3.19\times10^{-3}$Jy beam$^{-1}$.
Meanwhile, the changes between iterations are much smaller:
$\mathrm{RMS}(T_2-T_1)=9.22\times10^{-5}$Jy beam$^{-1}$ and $\mathrm{RMS}(T_3-T_2)=6.53\times10^{-5}$Jy beam$^{-1}$.
These results show that $T_1$ already approaches the converged solution closely, and the additional improvement from further iterations is at the sub-percent level relative to the first correction.
Figure 7 displays the spectral differences. In this frequency range, after the third iteration of stray radiation correction, $T_3$ showed a few percent increase from $T_a$, while $T_1$ differed slightly from $T_3$. According to the theory and results, the first iteration accounts for over 90\% of the flux change toward convergence, making the initial iteration of stray radiation correction sufficient for most cases.

\begin{center}
\includegraphics[width=16.0cm]{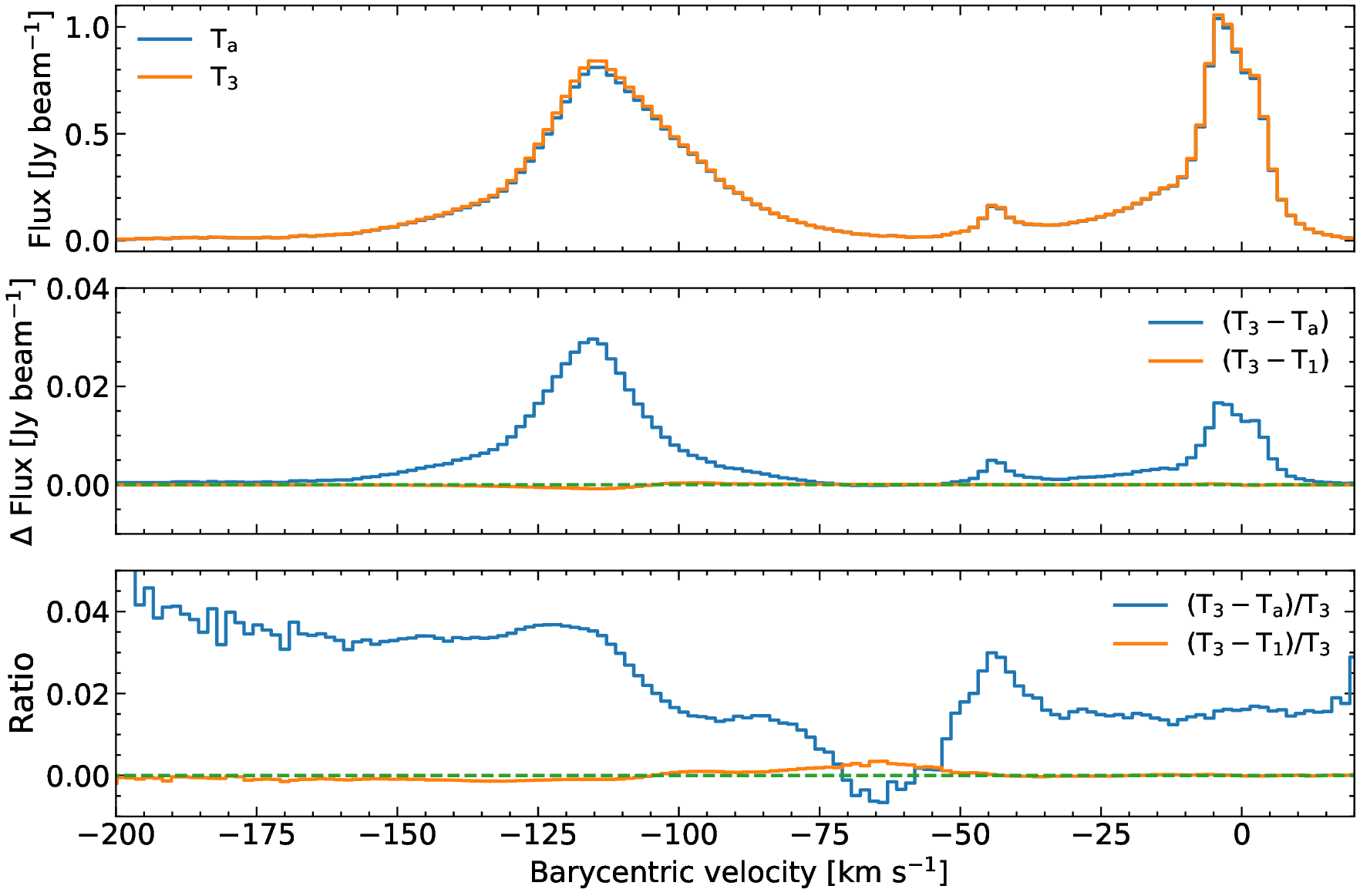}\\[8pt]
\parbox[c]{15.0cm}{\label{fig:freq}\footnotesize{\bf Fig.~7.}   The difference between the original spectrum and
the spectrum underwent one-time correction and three times iteration correction.
Upper: The spectrum between $-200~\kms$ and $20~\kms $. The emission of M33 (left peak) and the Milky Way (right peak) are included.
{Middle}: Differences between $T_a$, $T_1$, and $T_3$. The blue line represents $T_3-T_a$, the orange line for $T_3-T_1$, and the green dash line for zero.
{Lower}: The ratio of the difference between $T_3$ and $T_i$ to $T_3$. The blue line stands for 
$(T_3-T_a)/T_3$ and the orange line stands for $(T_3-T_1)/T_3$.
$T_2$ is not shown because it is nearly indistinguishable from $T_3$.
}

\end{center}

\section{Conclusions and Discussion} \label{sec:CD}

Stray radiation from radio telescopes can be received through both the main beam and the sidelobes. 
When a telescope is pointed away from a source, that source may still align with the telescope's sidelobes. Consequently, the flux from the source is incorrectly recorded as coming from the main beam’s current (off-source) position. This leads to an underestimation of the total flux when the telescope actually targets the source, as that 'leaked' flux is lost from the measurement. 
Despite the initial design of FAST to minimise this issue, approximately 2\% to 7\% of the total flux will still be spoilt into sidelobes, according to our observations of strong radio sources. This fraction depends on the configuration of beams in the L-band 19-beam receiver and rises as the beam moves away from the central beam. 

We carry out the correction for stray radiation on a per-beam basis due to the varying beam patterns and $\eta_{mb}$ values. To evaluate the impact of this correction, we compare the measurements before and after the adjustment. For extended sources, flux levels will exhibit fluctuations at different locations after correction.

By applying this method to FAST, we found that the approach is highly effective when the main-beam efficiency is high. In practice, a single iteration typically suffices for convergence, as demonstrated by our results. This behavior highlights the importance of high main-beam efficiency and suggests that future radio telescopes may benefit from adopting receiver designed with high main-beam efficiency.

Beyond its performance on FAST, the stray-radiation correction module, together with the HIFAST processing module, is open source. The HIFAST package is designed with a modular structure and can be ported to other radio telescopes that employ similar observing modes and data formats. Therefore, the strategy presented here is not limited to FAST, but can be transplanted and adapted to other facilities with appropriate beam models and survey data.

\section*{Data availability statement}
The beam patterns, consisting of 19 beams covering two frequency ranges: 1050–1140 MHz and 1330–1420 MHz, could be accessed on the stray radiation page of \texttt{the hifast} website\footnote{https://hifast.readthedocs.io/en/v1.3/hifast.sr.html} 
and Science Data Bank \footnote{https://www.scidb.cn/s/bqQRNv}.

\addcontentsline{toc}{chapter}{Acknowledgment}
\section*{Acknowledgment}
This work was supported by the China National Key Programme for Science and Technology Research and Development of China (2022YFA1602901, 2023YFA1608204), the National SKA Programme of China (No. 2022SKA0110201), 
the National Natural Science Foundation of China (NSFC, grant Nos. 11873051, 11988101, 12033008, 12041305, 12125302, 12173016, and 12203065), the CAS Project for Young Scientists in Basic Research grant (No. YSBR-062), the K.C. Wong Education Foundation, and the scientific research grants from the China Manned Space Project. 
Y.J. acknowledges support from the Cultivation Project for FAST Scientific Payoff and Research Achievement of CAMS-CAS. 

This work made use of the data from FAST (Five-hundred-metre Aperture Spherical radio Telescope) (\url{https://cstr.cn/31116.02.FAST}). FAST is a Chinese national mega-science facility operated by the National Astronomical Observatories, Chinese Academy of Sciences.



\end{document}